\documentclass[10pt,journal,a4paper,final,oneside,twocolumn]{IEEEtran}
\usepackage{amssymb,amsfonts,amsthm,amsmath,bm,color}
\usepackage{times}
\usepackage{graphicx}
\usepackage{cite}
\usepackage{color}
\usepackage{psfrag}
\usepackage{subfigure}
\usepackage{epsfig}
\usepackage{mathtools}
\usepackage{enumitem}
\usepackage{pifont}
\usepackage{array}
\usepackage{multicol}
\usepackage[T1]{fontenc}
\usepackage{comment}
\usepackage{acronym}
\usepackage{multirow}
\usepackage{epsfig}
\usepackage{hyperref}
\usepackage{breakurl}
\usepackage{setspace}
\theoremstyle{definition}
\newtheorem{thm}{Theorem}
\newtheorem{lem}{Lemma}
\def\h{{\mathbf h}}

\def\e{{\mathbf e}}

\def\0{{\mathbf 0}}

\def\I{{\mathcal I}}
\def\i1{{\mathbf 1}}
\def\Eb{{\mathbf E}}

\def\CN{{\mathcal{CN}}}

\def\E{{\mathbb E}}

\renewcommand{\baselinestretch}{1}
\usepackage[top=0.6in, bottom=0.6in, left=0.5in, right=0.5in]{geometry}
\begin{document}

\title{Cooperative 3D Beamforming for Small-Cell and Cell-Free 6G Systems}
\author{Sarath Gopi, Sheetal Kalyani, and Lajos Hanzo, {\em Fellow, IEEE}
%\thanks{The authors are with School of Electronics and Computer Science, University of Southampton, SO17 1BJ, UK. (E-mail: lx1g15, yl6g15, rm, lly, lh@soton.ac.uk). \emph{(Corresponding author: Yusha Liu)}}
%\thanks{L. Hanzo would like to gratefully acknowledge the financial support of the Engineering and Physical Sciences Research Council projects EP/Noo4558/1, EP/P034284/1, COALESCE, of the Royal Society's Global Challenges Research Fund Grant as well as of the European Research Council's Advanced Fellow Grant QuantCom (Grant No. 789028).}
}
\maketitle

	\begin{abstract}
	Three dimensional (3D) resource reuse  is an important design requirement for the prospective sixth generation (6G) wireless communication systems. Hence, we propose a cooperative 3D beamformer for use in 3D space. Explicitly, we harness multiple base station antennas for  joint zero forcing transmit pre-coding for beaming the transmit signals in specific 3D directions. The technique advocated is judiciously configured for use in both cell-based and cell-free wireless architectures. We evaluate the performance of the proposed scheme using the metric of Volumetric Spectral Efficiency (VSE). We also characterized the performance of the scheme in terms of its spectral efficiency (SE) and Bit Error Rate (BER) through extensive simulation studies. %These studies quantified the advantage of cell-free architectures over small-cells. 
\end{abstract}

\section{Introduction}
 
Industrial as well as academic research groups and standard bodies have embarked on studying the technical requirements of 6G systems. In contrast to 5G, 6G is expected to connect ``intelligence", rather than just ``things", while satisfying more stringent energy efficiency, latency and rate targets. Diverse technologies, such as Terahertz solutions, cell free massive multiple input multiple output (MIMO), 3D spatial modulation as well as the Internet of ``nano and bio things",  have been recommended for use in 6G systems \cite{letaief2019roadmap,rappaport2019wireless, akyildiz20206g, alsharif2020sixth,dang2020should}. 

\par 
Given the recent advances of non-terrestrial communication relying on unmanned aerial vehicles (UAVs) such as drones, airships and balloons, 6G is expected to harness them for enhancing both the coverage quality and capacity \cite{bucaille2013rapidly, mozaffari2019tutorial, mozaffari2018beyond, li2018uav}. Explicitly, when mmWave and Terahertz  frequencies are used, only Line-of-Sight (LOS) communication is  feasible owing to the high probability of 	blockage. Hence, the assistance of UAVs will become inevitable to provide services to users, where conventional terrestrial network is unavailable. These UAVs may play the role of both user equipment (UE)  and of aerial base stations (BS). These air-born UEs coexist with terrestrial ground users and are often 	termed as cellular-connected UAVs, which have been harnessed in applications such as remote sensing~\cite{xiang2019mini,partsinevelos2020novel}, package delivery~\cite{bamburry2015drones}, surveillance \cite{hosseinalipour2020energy} etc. A direct consequence of this new paradigm is that the UEs and BSs are going to co-exist in 3D space, where many of these have to engage in simultaneous communications. 
\par

Therefore, in 6G systems, the provision of services for aerial users at different altitude is also expected to become a requirement. In this context,  both the volumetric spectral-efficiency and energy-efficiency become important performance metrics. Hence, the 3D reuse of communication resources is expected to become an important differentiator between 5G and 6G~\cite{saad2019vision}. In support of the associated 3D resource allocation, we propose a novel 3D beamforming technique, which is capable of beneficially focussing the transmit signal in 3D space so that the spectral resources can be efficiently reused without excessive interference. {{Even though the authors of \cite{alam2006coverage, mozaffari2018beyond} and a few other references do indeed allude to 3D cellular structure, they do not recommend specific solutions for transmission to two or more users, who have the same angular locations with respect to a base station antenna using the same resources such as frequency, time, code etc. Our proposal is the first step towards this.  Explicitly, this is a transmit precoding technique, which may be combined with other techniques envisaged for meeting the challenging requirements of 6G systems \cite{chen2020massive}.}}
\begin{table*}[t]
	\begin{center}
		\begin{tabular}{|l|l|l|l|l|l|l|l|}
			\hline
			& Our Scheme & \cite{mozaffari2018beyond} & \cite{8770245} & \cite{afana2014performance} & \cite{xiong2020research}  &  \cite{ngo2017cell}   \\
			\hline
			3D Model &$\checkmark$ &$\checkmark$ &  & &$\checkmark$ &  \\
			\hline
			3D Beamforming &  $\checkmark$  &  &  $\checkmark$ & $\checkmark$ & &  \\
			\hline
			%3D Spatial & & & & & &  \\
			3D Spatial Utilization &  $\checkmark$ & &  & & &  \\
			\hline 
			%	Channel unaware & & & & & &  \\
			Channel unaware precoding &$\checkmark$ & &  & & &   \\
			\hline 
			%Cellular (CL) & & & & & & \\
			Cellular (CL)/Cell-free & Both & CL& CL  &  & CL & Both  \\
			\hline
			%	Rician  & & & & & &  \\
			Rician	channel &$\checkmark$ & & & & &   \\
			\hline
			Theoretical & Volume Spectral & &   &SINR, & &  \\
			expressions & Efficiency (VSE)   & Latency &  & BER & VSE &  \\
			\hline
		\end{tabular}
		\caption{Contrasting the proposed scheme to the existing solutions.}
	\end{center}
	\label{p6CompTable}
\end{table*}
\par
\begin{figure}
	\centering
	\includegraphics[width = 0.50\textwidth]{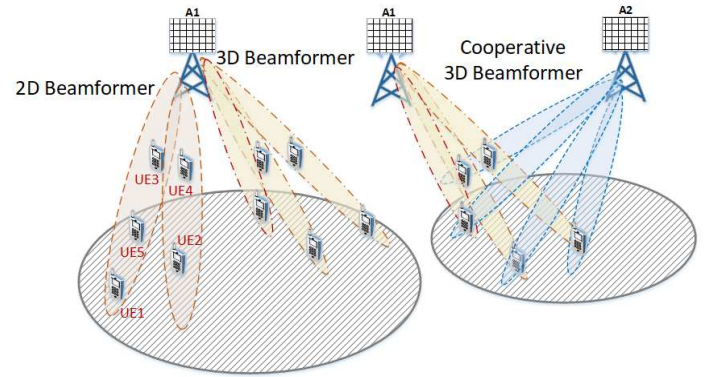}
	\caption{Difference between conventional 2D \& 3D beamformer and the proposed cooperative 3D beamformer.}	
	\label{p6figm1}
\end{figure}

 
{{A typical communication scenario showing the differences between the proposed beamforming techniques and the conventional 2D and 3D beamforming is portrayed in Fig~\ref{p6figm1}, where:}}
\begin{itemize}
	\item {{The UEs $UE1$, $UE2$ and $UE5$ are terrestrial users, while $UE3$ and $UE4$ are aerial users.}}
	\item {{With respect to the antenna array (AA) $A1$, the UEs $UE1$, $UE3$ and $UE5$ have the same azimuth angle, while $UE2$ and $UE4$ are located at a different, but between them identical azimuth angle. Hence, even a conventional 2D beamformer will be able to distinguish $UE1$, $UE3$ and $UE5$ from $UE2$ and $UE4$.} }
	\item {{However, simultaneous communication with $UE1$, $UE3$ and $UE5$ is not possible using a 2D beamformer  based precoder. Similar statement is valid for $UE2$ and $UE4$.}}
	\item {{$UE1$ and $UE3$ have the same elevation angle with respect to AA $A1$, which is however different from that of $UE5$. Hence, a 3D beamformer~\cite{razavizadeh2014three,chin2014emerging,8770245, 7146133}, which jointly uses the elevation angle along with the azimuth angle for precoding is indeed capable of distinguishing $UE5$ from $UE1$ and $UE3$. }}
	\item {{However, the pairs $UE1$ and $UE3$, and  $UE2$ and $UE4$ cannot be differentiated even using only 3D beamformer, since they have the same elevation and azimuth angles with respect to AA $A1$.}}
\end{itemize}
{{Hence, we propose cooperative bamforming, which exploits a pair of AAs, namely $A1$ and $A2$ seen at the right of Fig.~\ref{p6figm1} so that each fraction of the 3D space can be uniquely targeted. Therefore, all five UEs of Fig.~\ref{p6figm1} can simultaneously communicate using cooperative 3D beamforming.}}

%It can be seen that the proposed scheme is able to serve multiple users simultaneously, who are in the same 3D cone, where 3D beamformer fails to do so.

\par 
In addition to the UAV assisted scenario of Fig.\ref{p6figm1}, the proposed scheme may also be readily used for supporting Internet of Things (IoT). The evolution of IoT extended the application of wireless communications from smartphones and tablets to smart wearable, live health monitoring systems and diverse other sensors, which have new applications in agriculture, water supply and the smart grid \cite{zanella2014internet, gubbi2013internet, dawy2016toward, lien2011toward}. These applications support users in 3D space, especially in dense multi-story buildings of an urban scenario, while requiring a high data rate and ultra-low latency.  Hence, our cooperative 3D beamforming technique becomes beneficial in such a scenario.

\par 
A compelling technology for next generation wireless communication systems is constituted by massive multiple-input multiple-output (Massive MIMO) systems, where both the base station and the users have numerous antennas \cite{lu2014overview,gao2015massive},  hence exhibiting much improved throughput, energy efficiency, and reduced interference \cite{larsson2014massive}. We can have either co-located or distributed massive MIMO architectures. It has been shown in \cite{zhou2003distributed} that a distributed MIMO scheme achieves better coverage quality than its co-located counterpart owing to its improved diversity gain. A distributed architecture, where the antennas are geographically separated, acts as a MIMO network and it is also termed as a virtual MIMO or distributed antenna system (DAS) \cite{karakayali2006network,choi2007downlink,irmer2011coordinated,hong2013energy,castanheira2010distributed}. For exploiting the advantages of distributed systems, a cell-free architecture has been proposed in \cite{ngo2017cell, demir2021foundations} for next generation wireless communication, because it has been shown to outperform the maturing small-cell systems. To weigh up the associated pros and cons, we consider our proposed scheme in both small-cell and cell-free architectures. 

\par 
Our main contributions are: 
\begin{itemize}
	\item We propose a cooperative 3D beamforming (CBF) technique, relying on position based transmit beamforming (PBTBF). {{ As demonstrated in Fig.~\ref{p6figm1}, this is quite different from the conventional 3D cooperative beamforming, where a single AA is used for forming a sharp transmit beam in some specific azimuth and elevation angle pair \cite{afana2014performance, jayaprakasam2017distributed, 5449675, 4411470}.}} We may also refer to our cooperative 3D beamforming  scheme as joint transmit precoding (TPC) relying on a pair of AAs  to focus the transmit signal into the specified 3D region and thereby to assist in simultaneously serving the user in any 3D space, which is one of the 6G design requirements. Since 6G itself is in its planning phase, we believe that our proposal is the first of its kind. 
	\item {{ We exploit the metric termed as the average Volumetric Spectral Efficiency (VSE) for quantifying the performance of the proposed scheme. Explicitly, the VSE is defined as the sum of the maximum average number of bps/Hz/$km^3$ \cite{xiong2020research}.}} This is the 3D dual pair of the average Area Spectral Efficiency (ASE) metric of Alouni and Goldsmith \cite{Alouini1999}, which has been widely used as a performance metric in terrestrial communication \cite{li2016success, ding2015will, xin2015area, zhang2014generalized}. Similarly to the 2D ASE of conventional cellular systems, the 3D VSE captures the trade-off between the frequency reuse factor and the resultant link quality offered to the users. Hence, it will assist in the optimal allocation of resources in 6G systems.
	\item The proposed CBF technique may rely on both collocated  AAs and on distributed antennas, hence it  is beneficial for both small-cell and cell-free architectures. We demonstrate how to make this 3D resource reuse possible in both cell-free as well as small-cell architectures and study the impact of these architectures in the context of VSE.
	\item We derive the mathematical expression for the VSE of the proposed scheme and compare the same with extensive simulation results. 
\end{itemize}
In Table \ref{p6CompTable}, we provide a bold summary and contrast our new contributions to the seminal literature. The proposed scheme can also be used both at mmWave and the Terahertz frequencies, hence it is beneficial for 6G systems. {{A significant advantage of our proposed scheme is that the transmitter does not have to rely on any channel knowledge for pre-coding, rather it simply relies on the user positions.}} In Section \ref{p6prop}, we develop the proposed CBF technique. The performance metric VSE is defined and derived in Section \ref{vse}. Our simulation results are detailed in Section \ref{p6sim}, and the paper is concluded in Section \ref{p6con}.

\section{Proposed Scheme}
\label{p6prop}
\begin{figure}
	\centering
	\includegraphics[width = 0.450\textwidth]{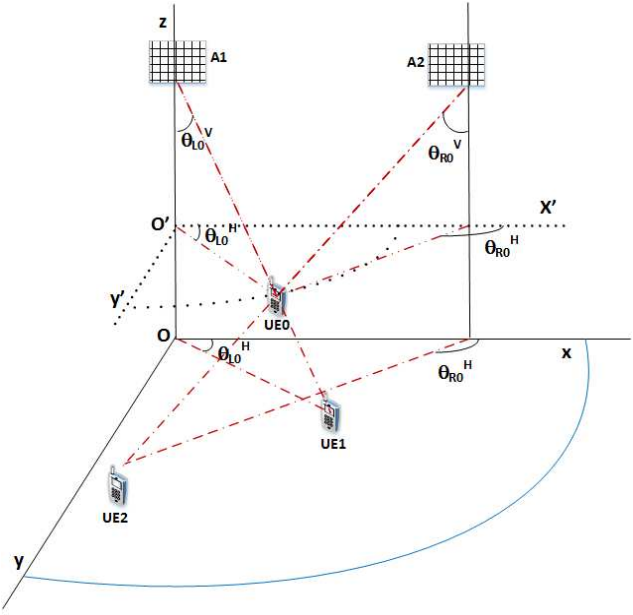}
	\caption{The 3D geometry considered.}	
	\label{p6fig0}
\end{figure}

The basic idea of the proposed scheme is discussed using Fig.~\ref{p6fig0}. Let us assume having a pair of AAs $A1$ and $A2$, each having $M$ elements, where $A1$ and $A2$ are separated in 3D space and the users are spatially distributed in 3D all around these AAs. It is assumed that each UE has a single antenna.  Let us assume that $UE0$ is the user of interest, located at the 3D angles of $\theta_{L0}$ and $\theta_{R0}$ with respect AAs $A1$ and $A2$, respectively. Here $\theta_{L0} = \left(\theta_{L0}^H, \theta_{L0}^V\right)$ is the azimuth-elevation angle pair of $A1$  and similarly $\theta_{R0}= \left(\theta_{R0}^H, \theta_{R0}^V\right)$  is that of $A2$. $UE0$ is shown to be an aerial user in Fig.~\ref{p6fig0} with $X'O'Y'$ being a horizontal plane parallel to the ground plane $XOY$. We consider the case, where there are many interferers and these can be located anywhere in space. However, we specifically consider the pair of interferers $UE1$ and $UE2$ in Fig.~\ref{p6fig0}. Here, $UE1$ has the same 3D angle as that of $UE0$, i.e.,  $\theta_{L0}$ with respect to $A1$, whereas the 3D angle of $UE2$ with respect to $A2$ is $\theta_{R0}$, which is the same as that of $UE0$. Hence, a conventional 3D beamformer using either $A1$ or $A2$ fails to simultaneously communicate with these three users. However, the interference emanating from users, who do not have the same 3D angle as that of the desired user can be readily eliminated using conventional 3D beamforming or zero-forcing TPC by a single AA \cite{Khan2010, Litva1996, Yoo2003,9491998}. For example, in Fig. \ref{p6fig0} transmission to both $UE1$ and $UE0$  can be arranged simultaneously using  conventional 3D beamforming at AA $A2$. However, our 3D CBF technique is required for simultaneous transmission to all of $UE0$, $UE1$ and $UE2$ in Fig.\ref{p6fig0}.
\par 
In order to elaborate further on the proposed 3D CBF scheme, we make the following assumptions:
\begin{enumerate}[label=a\arabic*)]
	\item Each AA knows the locations of the other AA and all users. Furthermore, they know the information to be transmitted to the user.
	\item Perfect synchronization exists between two AAs.
	\item Each user has the perfect knowledge of its channel spanning from both AAs. {{Note that this is a common requirement of any coherent communication system. Existing techniques may be adopted for channel estimation \cite{zhao2017multi}.}} However, note that only the user needs the channel information for decoding,  whereas the AAs only require location information for precoding.
	\item A strong line-of-sight (LOS) path exists between each and every user, hence we have Rician channels between the AA and the user.
\end{enumerate}
Note that the above assumptions may indeed be satisfied in practise. For example, a base station controlling the two antenna arrays can share information among them and synchronize them. A fibre link between the AAs can also be used to achieve this. In any communication system, the channel information is essential for detecting the symbols and can be acquired by channel estimation. However, in our case, the users have to estimate two channels, namely between themselves and the pair of AAs. Finally, it is widely recognized that a strong LOS path is necessary for reliable communication at mmWave and Terahertz frequencies \cite{liu2016line, rappaport2015wideband, rappaport2019wireless}. In the proposed scheme, we use antenna arrays to form beams towards the user and thereby rely on a LOS path. Let us now detail the proposed 3D CBF scheme.
%Therefore, we are more concerned with interferers, which happen to be in either of the cone of the two AAs supporting the desired user. For example, the pair of  interferers $UE1$ and $UE2$ seen in Fig.~\ref{p6fig10} are in the receiver cone of $UE0$ impinging from the AA $A1$ and that of $UE0$ emerging from AA $A2$, respectively.

\par 
Let $s_0$ be the desired signal to be transmitted to $UE0$. Let us assume furthermore that this signal is transmitted from both AAs $A1$ and $A2$ by the TPC vector $\e_0$ of dimension $M \times1$. {{The channel between $A1$ and $UE0$  is denoted by $\h_{L0}$ and that between $A2$ and $UE0$ as $\h_{R0}$. Note that the dimension of the channel vectors is $M \times1$.}} As stated in assumption (a4), both $\h_{L0}$ and $\h_{R0}$ are represented by Rician channels. Then the signal received at $UE0$ in the absence of any other user is given by:
\begin{align}
y_0 = \left(\h_{L0} +\h_{R0}\right)^H \e_0s_0+w_0.
\label{p6eqn1}
\end{align}
Let us assume that there are $N$ interferers, with $s_n, 1 \le n \le N$  being the symbols to be transmitted to the $n^{th}$ $UE$ and let $\e_n$  be the corresponding TPC vector. Hence, the received signal of $UE0$ is given by:
\begin{align}
y_0 = \left(\h_{L0} +\h_{R0}\right)^H \left(\sum_{n=0}^{N}\e_ns_n\right) +w_0,
\label{p6eqn2}
\end{align}
where $n=0$ corresponds to the terms related to the desired user and $w_0 \sim \CN(0,\sigma^2)$ is the AWGN component. Let $\e_{Ln}$ and $\e_{Rn}$ represent the position-based steering vectors of the $n^{th}$ UE with respect to AAs $A1$ and $A2$, respectively. Explicitly, $\e_{Ln}$ and $\e_{Rn}$ are $M \times  1$ vectors, whose $i^{th}$ entry is of the form $exp\left[j\frac{2\pi}{\lambda}\left(x_i\psi_n^x+y_i\psi_n^y+z_i\psi_n^z\right)\right]$. Here, for $\e_{Ln}$, we have $\psi_n^x=\sin \theta_{Ln}^V \cos \theta_{Ln}^H$, $\psi_n^y=\sin \theta_{Ln}^V \sin \theta_{Ln}^H$ and $\psi_n^z=\cos \theta_{Ln}^V$. Similarly, for $\e_{Rn}$, the corresponding values are $\psi_n^x=\sin \theta_{Rn}^V \cos \theta_{Rn}^H$, $\psi_n^y=\sin \theta_{Rn}^V \sin \theta_{Rn}^H$ and $\psi_n^z=\cos \theta_{Rn}^V$, where $(x_i,y_i,z_i)$ represents the corresponding coordinate of the $i^{th}$ element in $A1$ and $A2$ respectively, for $\e_{Ln}$ and $\e_{Rn}$. Therefore,  $\h_{L0}$ and $\h_{R0}$ can be written under the Rician channel model as \cite{xu2009approximation}:
\begin{align}
\h_{L0} = \sqrt{\frac{K}{K+1}}\e_{L0} + \sqrt{\frac{1}{K+1}} \h_L
\label{p6eqn3}
\end{align}
and
\begin{align}
\h_{R0} = \sqrt{\frac{K}{K+1}}\e_{R0} + \sqrt{\frac{1}{K+1}} \h_R,
\label{p6eqn4}
\end{align}
where the first term at the RHS of (\ref{p6eqn3}) and (\ref{p6eqn4}) is the LOS component, while $\h_L$ and $\h_R$ represent the nonLOS components. Furthermore it is assumed that each entry of $\h_L$ and $\h_R$ is an i.i.d zero mean complex Gaussian random variable. Let $\h_0 = \h_{L0}+\h_{R0}$. {{We need to find out a precoding vector such that users in each 3D locations are uniquely communicated from the BS. Explicitly, this is an interference cancellation problem, for which we are proposing the following different solutions.} }

\subsection{Proposed Solutions}
\label{prop}
\subsubsection{Zero Forcing Precoder (ZFP)}
In the zero forcing precoder, we propose to select the position-based joint TPC (JTPC) vector $\e_0$ as the solution of the following optimization problem.
\begin{align}
\underset{\e_0}{\min}~&\|\e_0 - \left(\e_{L0}+\e_{R0}\right)\|^2~ \nonumber \\
\text{such that}~&\e_0^H\left(\e_{Ln}+\e_{Rn}\right)=0 \nonumber \\
&~~~~~~\text{for each interferer}~n=1, 2, ..., N.
\label{p6eqn5}
\end{align}
The solution of (\ref{p6eqn5}) is \cite[pp. 169]{van2004optimum}: 
\begin{align}
\e_0 = \left(\I- \Eb_0 \left(\Eb_0^H\Eb_0\right)^{-1}\Eb_0^H \right)\left(\e_{L0}+\e_{R0}\right),
\label{p6eqn6}
\end{align}
where $\Eb_0$ is the matrix whose columns are $\left(\e_{Ln}+\e_{Rn}\right), n=1, ..., N$. Note that the solution of (\ref{p6eqn5}) is the zero-forcing JTPC vector for the space spanned by $\left(\e_{L0}+\e_{R0}\right)$, whose response is zero in the space spanned by $\left(\e_{Ln}+\e_{Rn}\right)$ for $n \ne 0$. Similarly, the JTPC vector for each user may also be readily computed. Note that the JTPC vector $\e_0$ of (\ref{p6eqn6}) completely cancels the contributions of all LOS components of the interferers.
\par 
{{Additional derivative constraints can be added to (\ref{p6eqn5}) in order to make the solution robust against position errors \cite[3.7.2]{van2004optimum}. Explicitly, we add the following first-order derivative constraints:
		\begin{align}
		\bar{\e}_n^a =\frac{\partial}{\partial \theta_n^a} \left(\e_{Ln}+\e_{Rn}\right) &= 0, ~n= 0, 1, ..., N\nonumber \\
		\bar{\e}_n^z =\frac{\partial}{\partial \theta_n^z} \left(\e_{Ln}+\e_{Rn}\right) &= 0, ~n= 0, 1, ..., N.
		\label{p5eqnn4}
		\end{align}
		%Note that the derivative constraints are added to the desired user's steering vector only. We can add similar constraints to the interfering users' steering vectors also. However, this may result in degrading the condition number of $\left(\Eb_0^H\Eb_0\right)$, if the number of antenna elements are not sufficient.
		The solution will have the same form as that of (\ref{p6eqn6}), except that $\Eb_0$ is replaced by $\hat \Eb_0$, where $\hat \Eb_0 = \Eb_0 \cup \{\bar{\e}_0^a ~\bar{\e}_0^z\}$. We name this solution as ZFP-D. Note that adding these additional constraints may result in degrading the condition number of $\left(\hat \Eb_0^H \hat \Eb_0\right)$, if the number of antenna elements is not sufficient. In such a case, the particular singular value decomposition (SVD) based scheme may be adopted to determine the inverse of $\left(\hat \Eb_0^H \hat \Eb_0\right)^{-1}$ \cite[pp. 423]{meyer2000matrix}.}}

\subsubsection{Minimum Power Distortionless Response (MPDR) Precoder}

{{Another precoding solution could be based on the MPDR beamformer, where the total output power is minimized subject to a distortionless constraint \cite{van2004optimum}. Mathematically, it is given as: }}

\begin{align}
\underset{\e_0}{\min} ~  \|\e_0^H  \tilde \Eb_0\|^2 ~\text{such that}~  \e_0^H\left(\e_{L0}+\e_{R0}\right)= \bm{1},
\label{p6eqnn7}
\end{align}

{{where $\tilde \Eb = \Eb_0 \cup \left(\e_{L0}+\e_{R0}\right)$. The precoding vector obtained by solving the optimization problem (\ref{p6eqnn7}) is \cite{van2004optimum}:}}
\begin{equation}
\e_0 = \frac{ \left[\tilde\Eb_0 \tilde \Eb_0^H\right]^{-1} \left(\e_{L0}+\e_{R0}\right)}{\left(\e_{L0}+\e_{R0}\right)^H \left[\tilde \Eb_0 \tilde \Eb_0^H\right]^{-1} \left(\e_{L0}+\e_{R0}\right)}.
\label{p6eqnn8}
\end{equation}
{{However, when the number of antenna elements and users results in an ill conditioned matrix $\left(\tilde \Eb_0 \tilde \Eb_0^H\right)$, then the method adopted in \cite[(13)]{9491998} may be used to get (\ref{p6eqnn8})}.}

\subsection{General Case}

{{ The channel model of (\ref{p6eqn3}) and (\ref{p6eqn4}) has the same Rician factor for both the channel impinging from first and second antenna arrays. However, it may indeed happen that they are not the same. In such a scenario, we have the following optimization problem:}}
\begin{align}
\underset{\e_0}{\min}~&\|\e_0 - \e_{L0}\|^2 + \|\e_0 - \e_{R0}\|^2~ \nonumber \\
\text{such that}~&\e_0^H\e_{Ln}=0,~n=1, 2, ..., N \nonumber \\
~&\e_0^H\e_{Rn}=0,~n=1, 2, ..., N.
\label{p6eqn9}
\end{align}
{{ The solution of (\ref{p6eqn9}) will have the same form as that of (\ref{p6eqn6}), except that $\Eb_0$ is replaced by $\bar{\Eb}_0$ given below:}}
\begin{align}
\bar{\Eb}_0 = \left[\e_{L1}~\e_{R1}~\e_{L2}~\e_{R2}~...~\e_{LN}~\e_{RN} \right].
\label{p6eqn9a}
\end{align}

{{Note that in (\ref{p6eqn9}), there are $(2N)$ constraints, whereas in our original problem (\ref{p6eqn5}), there are only $N$ constraints. Hence, when the number of interferers is high, (\ref{p6eqn9}) may require more elements in the AAs to find a feasible solution. Therefore, whenever the Rician factors in (\ref{p6eqn3}) and (\ref{p6eqn4}) are the same, it is better to use the formulation in (\ref{p6eqn5}).  Now in Section \ref{subsectsc} and \ref{subsectcf}, the specific small-cell and cell-free scenarios are discussed.}}

\subsection{Small-Cell Architecture}
\label{subsectsc}
In small-cell based systems, the AAs $A1$ and $A2$ of Fig.~\ref{p6fig0} can be a pair of cooperating BSs, each equipped with $M$-element AAs. However, the concept is not restricted to a twin BS configurations, it can also work for more complex BS configurations. For example,  a single BS having two antennas, which are sufficiently separated can also be used. Another option is to have a single transmit antenna cooperating with a pair of Reconfigurable Intelligent Surfaces (RIS) controlled by a BS. Explicitly, a RIS is an array of passive reflecting elements relying on low-complexity integrated electronics for introducing a beneficial phase shift to the incident waves \cite{qingqing2019towards,dai2019reconfigurable,di2019smart}. This has been advocated as a low-cost, energy-efficient solution for a diversity of applications \cite{yan2019passive,yu2019miso,han2019large,huang2019reconfigurable, huang2018energy,chen2019intelligent,pan2020multicell,basar2020reconfigurable,9226135} and can be used to provide the necessary phase shifts for the proposed scheme.
\par
\begin{figure}
	\centering
	\includegraphics[width = 0.25\textwidth]{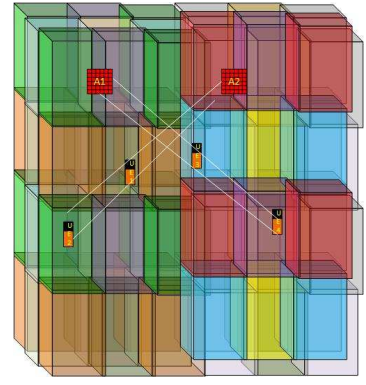}
	\caption{Proposed architecture for small-cell system.}	
	\label{p6fig1_1}
\end{figure}

 
Fig.~\ref{p6fig1_1} shows the example of a 3D cellular structure. The AAs $A1$ and $A2$ are intended to serve a large volume, which can be considered as a 3D macro-cell. The size of the macro-cell is limited by the maximum communication range of the AAs. This is partitioned into many small cells, termed as micro-cells, which are represented using different colours. The specific micro-cells having the same colours use the same resources. Similar to the case of existing cellular networks, the resources are shared by ensuring the same spectral resources are not used in the nearby micro-cells.  In Fig.~\ref{p6fig1_1}, the micro-cells are of cuboid shape, but can be of different shapes, such as hexagonal prisms, truncated octahedrons,  rhombic dodecahedrons and so on \cite{alam2006coverage}. The truncated octahedron is the closest approximation of a sphere and the number of polygons required for completely filling the 3D space is known to be the lowest for this particular structure \cite{alam2006coverage, mozaffari2018beyond}.  Finally, as detailed in Section \ref{p6prop}, by beneficially adjusting the phase shifts in $A1$ and $A2$, any user in the macro-cell can be served with minimal interference from other users exploiting the same resources. For example, the users $UE1$ and $UE2$ seen in Fig.~\ref{p6fig1_1}, who have the same 3D angle as that of the AA $A1$ can be simultaneously served. Similar is the case of $UE3$ and $UE4$, which are located in the same cone of AA $A2$.
\subsection{Cell-Free Architecture}
\label{subsectcf}
\begin{figure}
	\centering
	\includegraphics[width = 0.40\textwidth]{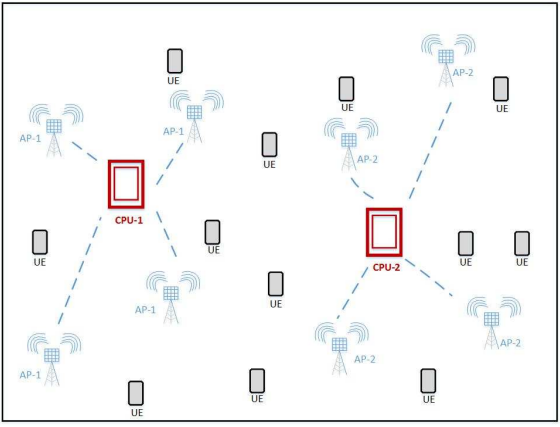}
	\caption{Proposed architecture for cell-free system.}	
	\label{p6fig1_2}
\end{figure}

We model our cell-free architecture as shown in Fig. \ref{p6fig1_2}, which is based on the one proposed in \cite{ngo2017cell} subject to some changes. In contrast to the small-cell based architecture, here the elements of the antenna arrays $A1$ and $A2$ are distributed throughout the cells. Specifically, the access points (APs) in Fig. \ref{p6fig1_2} act as the elements in $A1$ and $A2$. In contrast to \cite{ngo2017cell}, the proposed cell-free architecture has the following differences. 
\begin{itemize}
	\item There are two Central Processing Units (CPU), one supporting each of the distributed antenna arrays $A1$ and $A2$. 
	\item  The antenna at the APs has more than one element. In principle, the spacing between the elements in an AA should be arranged to be less than or equal to half the wavelength of the maximum operating frequency. This condition avoids grating lobes, hence pointing the signal in the desired direction~\cite{van2004optimum}. Since the APs host the geometrically separated elements of AAs, their half-wavelength spacing cannot be ensured in the distributed architecture. Hence, it is proposed to use a small AA in each of the APs, where the spacing is fixed to less than or equal to half the wavelength. Hence, in the cell-free architecture, the AAs $A1$ and $A2$ circumvent the grating lobe problem.
	
\end{itemize}

\par 
The entire set of APs is partitioned into two groups, each of which is associated with one of the CPUs. Note that the number of APs associated with each CPU can be the same or different. The angle pairs $\theta_L$ and $\theta_R$ of the users are defined with respect to the two CPUs. In addition to the users' location, each AP should know the location of its respective CPU. The phase shifts to be applied at the elements of the AP to the transmit signal are calculated based on the relative positions of the corresponding CPU and the users.  The basic principle is the same as that in the small-cell based architecture, except that the distributed APs act as AA $A1$ and $A2$. {{Note that the flexible selection of two CPUs can be dynamically updated in such a way that the desired user is in the cell-centre, which is one of the key features of cell-free systems. Methods suggested for ensuring this in conventional cell-free applications may be appropriately modified for the proposed scenario, which may form part of our future work.}}
\par 
The cell-free architecture is more suitable for UAV specific applications, since there may be numerous UAVs spread across the volume considered, some of which can act as APs. Now, in Section \ref{vse}, we will first define and then deriving the Volumetric Spectral Efficiency (VSE) of the proposed scheme. 
\section{Volumetric Spectral Efficiency}
\label{vse}
We define the average VSE as:
\begin{align}
VSE = \frac{\bar{C}}{V},
%V_e = \frac{\bar{C}}{\frac{4}{3}\pi \left(d/2\right)^3},
\label{p6eqn7}
\end{align}
where $\bar{C} = \E\{\log \left(1+\eta\right)\}$ is  the average normalized channel capacity per Hertz with $\eta$ being the Signal to Interference plus Noise Ratio (SINR) and $V$ is the volume. Note that (\ref{p6eqn7}) is similar to the ASE equation \cite[(8)]{Alouini1999}, except that the denominator is the volume instead of the area.
\par 
Recall that the conventional terrestrial cellular structure tends to rely on perfectly tessellated hexagonal cells, which are often approximated by circles. However, in our case, the base station has to serve users in space. Hence,  we propose to extend the circles to the third dimension so that the resultant spherical cellular structure covers the entire 3D user space. Hence, in this case, a BS serves the users spanned in a sphere. This can be considered as a macro-cell. Now, we partition each macro-cell into small spherical micro-cells, where the users in a micro-cell are provided with orthogonal resources. However, the same resource can be shared by the users in different micro-cells of the same macro-cell. This makes it possible to reuse the resources in a macro-cell.  Let $d$ be the diameter of the micro-cell and $N_R$ be the reuse factor in the small-cell structure. Then the volume $V$ is: 
\begin{align}
V = N_R\frac{4}{3}\pi \left(d/2\right)^3.
\label{p6eqn7_1}
\end{align}
\par 
Note that for cell-free architectures, the denominator in (\ref{p6eqn7}) should be the volume of the region, where the average capacity $\bar{C}$ is estimated divided by the number of users sharing the same resources in that region. Assuming unity power for the transmitted symbols, the instantaneous  SINR  can be written from (\ref{p6eqn1}) as:
\begin{align}
\eta = \frac{P_r\|\h_0^H\e_0s_0\|^2}{\|\sqrt{P_r}\h_0^H\sum_{n=1}^{N}\e_n s_n+ w_0\|^2},
\label{p6eqn8}
\end{align}
where $P_r$ is the received signal power. We now derive the average channel capacity of the system in Theorem \ref{p6thm1} given below.
\begin{thm}
	\label{p6thm1}
	Given that the channel statistics and the desired as well as interfering users' locations are fixed, the average channel capacity per unit bandwidth for the zero forcing precoding is:
	\begin{align}
	&\bar{C}_{h,u}  = \frac{1}{\log 2} \left(\log \left(c_{\sigma}k_s^2\right) - Ei(k_s^2) + \zeta \right. \nonumber \\
	&~~~ \left.+ \frac{e^{-\frac{c_{\sigma}k_s^2}{c_{\sigma}-1}}}{c_{\sigma}-1}\left(Ei\left(\frac{c_{\sigma}k_s^2}{c_{\sigma}-1}\right)- Ei\left(\frac{k_s^2}{c_{\sigma}-1}\right) \right) \right)%.\nonumber \\
%	&~~~~~~\left.-\frac{2 }{k_s^2 } \left(1 - e^{-k_s^2}\right)\right), 
	\label{p6eqn15e}
	\end{align}
	%	\begin{align}
	%\bar{C}_{h,u}  &=   \frac{1}{ \left(c_{\sigma}-1\right) k_s^2\log 2}   e^{\frac{c_{\sigma}k_s^2}{1-c_{\sigma}}}\left(Ei\left[\frac{k_s^2}{c_{\sigma}-1}\right]-Ei\left[\frac{c_{\sigma}k_s^2}{c_{\sigma}-1}\right]\right)  \nonumber \\
	%&~~+ \frac{1}{k_s^2\log 2}\left(\gamma k_s^2 - Ei\left[-k_s^2\right] +\log \left(c_{\sigma} k_s^2\right) \right. \nonumber \\
	%&~~~~~~\left.-(1+k_s^2) \left(Chi(k_s^2)-Shi(k_s^2)\right) \right), 
	%\label{p6eqn15e}
	%\end{align}
	where $k_s = \frac{|\mu_s|}{\sigma_s}$ is the ratio of the absolute value of the average signal amplitude and its standard deviation, while $c_{\sigma}= \frac{\sigma_s^2}{\sigma_I^2}$ is the ratio of the variances of the signal and of the interference plus noise components. Furthermore, in (\ref{p6eqn15e}) are $Ei[x]=-\int_{x}^{\infty} \frac{e^{-t}}{t}dt$ represents the Exponential integral with $\zeta \approx 0.577$ being the Euler-Mascheroni constant~\cite[p. 278]{l1994concrete}.
	
	\begin{proof}
		See Appendix \ref{p6proofthm1} for proof.
	\end{proof}
\end{thm}
{{Now, we will consider the case, where the transmitted symbols are chosen from an equally likely zero mean constellation. In this case, the channel capacity per unit bandwidth is given in Theorem \ref{p6thm2} below.}}
\begin{thm}
	\label{p6thm2}
	{{When the transmitted symbols are chosen from an equally probable and zero mean constellation, the average channel capacity per unit bandwidth in (\ref{p6eqn15e}) can be written as:}}
	\begin{align}
	\bar{C}_{h,u}  =   \frac{c_{\sigma}}{c_{\sigma}-1}\log_2 c_{\sigma} 
	\label{p6eqn15f}
	\end{align}
	\begin{proof}
		See Appendix \ref{p6proofthm2} for proof.
	\end{proof}
\end{thm}
\par
{{In a wide range of applications, the transmitted symbols are chosen from an $M$-ary QAM constellation, where the conditions of Theorem \ref{p6thm2} are satisfied and hence (\ref{p6eqn15f}) can be used for estimating $\bar{C}_{h,u}$.}} Finally, the average capacity per unit bandwidth can be found by integrating (\ref{p6eqn15e}) or  (\ref{p6eqn15f}) over all possible user and interference locations ($u$) and is formulated as: 
\begin{align}
%\bar{C} = \int \left(\int_{0}^{\infty} \int_{-\infty}^{\infty}  \bar{C}_{h,u} f_{h_L}(h_L) f_{h_R}(h_R)dh_Ldh_R \right)f_u(u)du,
\bar{C} = \int  \bar{C}_{h,u} f_u(u)du,
\label{p6eqn16}
\end{align}
{{where $f_u(u)$ is the distribution of the desired and interfering users' location and $\bar{C}_{h,u}$ is given by either (\ref{p6eqn15e}) or (\ref{p6eqn15f}), which is a function of $c_{\sigma}= \frac{\sigma_s^2}{\sigma_I^2}$ with $\sigma_s^2$ and $\sigma_I^2$ are given by (\ref{p6eqn20}) and (\ref{p6eqn21}), respectively. Hence, $\bar{C}_{h,u}$ is a function of the precoding vector $\e_n~\forall~n=0, 1, ..., N$, which inherently depends on the desired user's and interferers' locations through the steering vectors $\e_{Ln}$ and $ \e_{Rn}$.}} Therefore, (\ref{p6eqn16}) represents a multi-dimensional integral over all the possible desired and interfering user locations. The integral in (\ref{p6eqn16}) cannot be solved analytically, but it can be evaluated by using Monte Carlo simulations. Finally, substituting (\ref{p6eqn16}) into (\ref{p6eqn7}) will give the average VSE of the proposed scheme. 
\par 
{{Note that $\bar{C}$ increases with $c_{\sigma}$, which is a measure of the signal to interference plus noise ratio. It becomes clear from (\ref{p6eqn20}) and (\ref{p6eqn21}) that for each desired user, $c_{\sigma}$ increases as a function of its precoding vector magnitude of $0^{th}$ user, i.e., $|\e_0|$. However, this will reduce the SINR for another user. Hence, it is best to have equal strength for each users' precoding vector and this will be maximum, if the steering vectors for each users are orthogonal to each other. The grade of orthogonality between the steering vectors can be readily increased by incorporating more elements into the transmit AAs. Similarly, given the number of elements in the AAs and a specific spatial volume, there is a particular maximum number of users, who can have orthogonal steering vectors. Hence, increasing the number of users to these values will increase the channel's utility. However, any further increase in the number of users may erode the capacity.}} In Section \ref{p6sim}, the VSE obtained through simulations is compared against the theoretical VSE of (\ref{p6eqn7}).
\section{Simulation Results}
\label{p6sim}

\begin{figure}
	\centering
	\includegraphics[width = 0.45\textwidth]{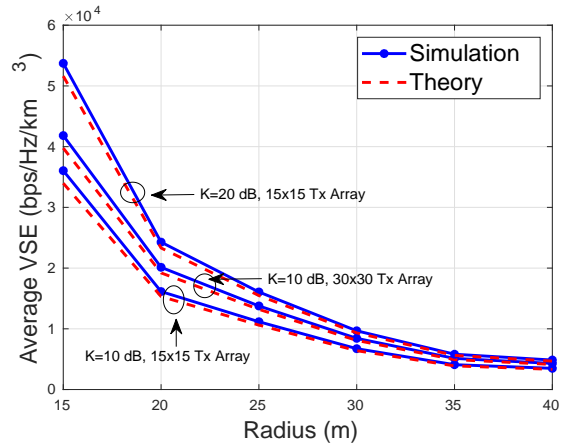}
	\caption{Average VSE vs Cell Radius.}	
	\label{p6fig2}
\end{figure}

\begin{figure}
	\centering
	\includegraphics[width = 0.45\textwidth]{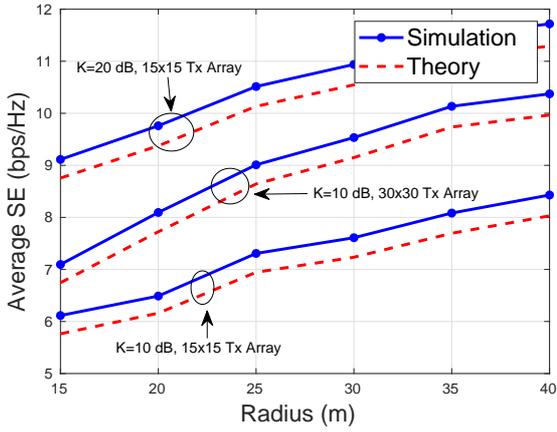}
	\caption{Average SE vs Cell Radius.}	
	\label{p6fig3}
\end{figure}

\begin{figure}
	\centering
	\includegraphics[width = 0.45\textwidth]{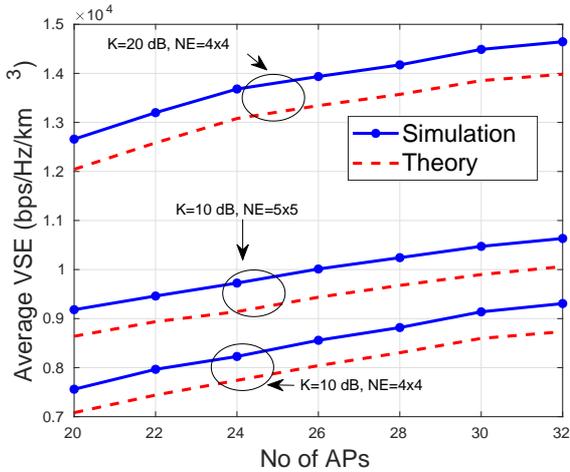}
	\caption{Average VSE vs No of APs.}	
	\label{p6fig5}
\end{figure}

\begin{figure}
	\centering
	\includegraphics[width = 0.45\textwidth]{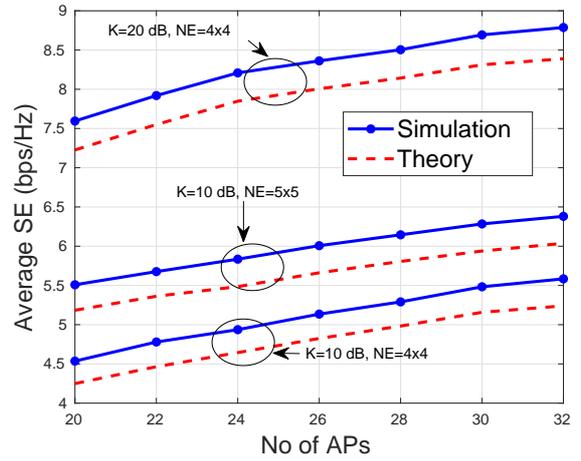}
	\caption{Average SE vs No of APs.}	
	\label{p6fig6}
\end{figure}

\begin{figure}
	\centering
	\includegraphics[width = 0.45\textwidth]{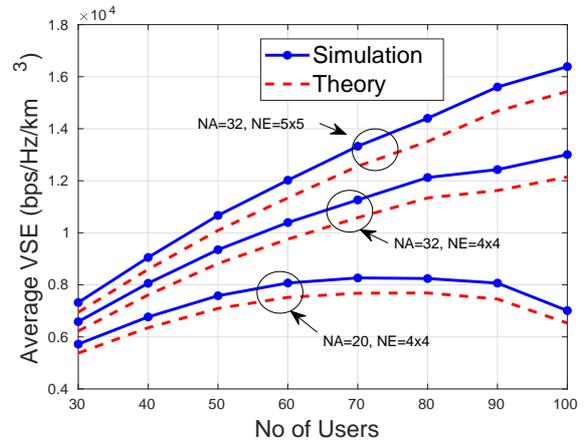}
	\caption{Average VSE vs No of Users.}	
	\label{p6fig8}
\end{figure}

\begin{figure}
	\centering
	\includegraphics[width = 0.45\textwidth]{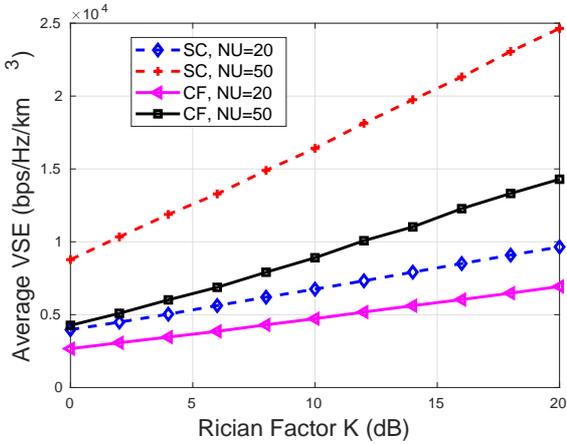}
	\caption{Average VSE vs Rician Factor K (dB). $SC$ and $CF$ represent small-cell and cell-free configurations, respectively.}	
	\label{p6fig10}
\end{figure}

\begin{figure}
	\centering
	\includegraphics[width = 0.45\textwidth]{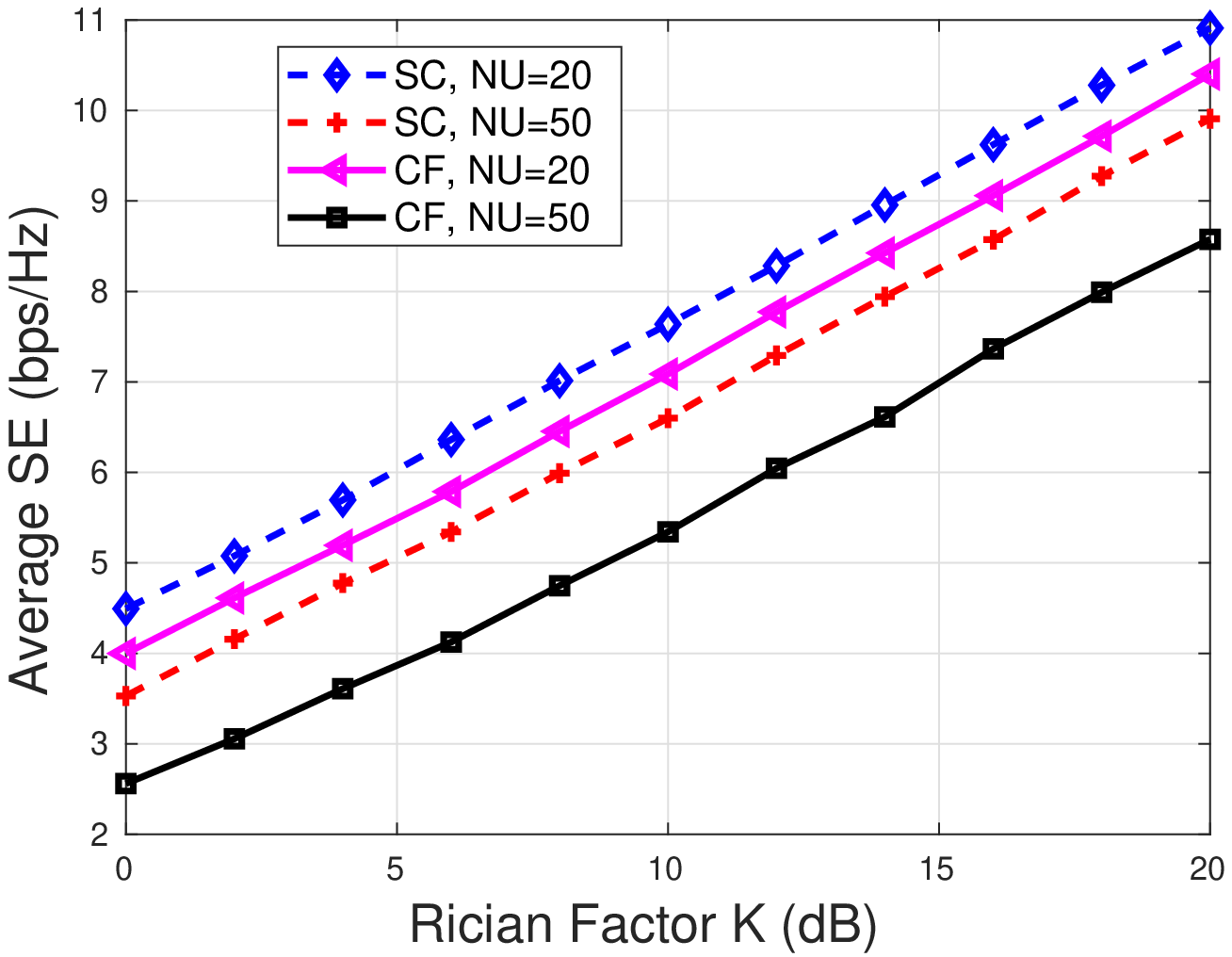}
	\caption{Average SE vs Rician Factor K (dB). $SC$ and $CF$ represent small-cell and cell-free configurations, respectively.}	
	\label{p6fig11}
\end{figure}

\begin{figure}
	\centering
	\includegraphics[width = 0.45\textwidth]{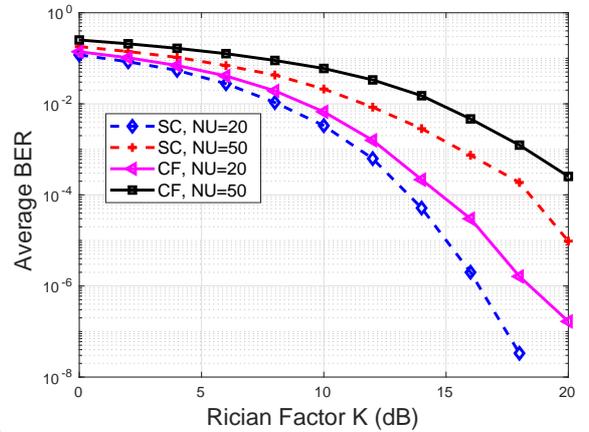}
	\caption{BER vs Rician Factor K (dB). $SC$ and $CF$ represent small-cell and cell-free configurations, respectively.}	
	\label{p6fig12}
\end{figure}

The proposed 3D CBF technique has been evaluated through extensive simulation studies. We characterized these techniques in terms of its average VSE in $bps/Hz/km^3$, its spectral efficiency (SE) in $bps/Hz$ and its Bit Error Rate (BER). Even though the VSE and SE are computed from the same quantity, they have different implications. For example, for a given number of users, the SE is increased as the volume is increased due to the reduction of the interference, while the VSE may increase or decrease. The VSE may be conveniently used for determining the optimal number of users in a given volume or the optimal volume for a given number of users. On the other hand, the SE quantifies the maximum total information rate in a given scenario. Therefore, the SE is useful for comparing two schemes, once the frequency-reuse topology is fixed, whereas the VSE is critical for determining the resource-reuse topology. Fig.~\ref{p6fig2} and Fig. ~\ref{p6fig3} show the results of the proposed scheme for our small-cell based architecture, whereas Fig.~\ref{p6fig5}-Fig.~\ref{p6fig8} characterize the cell-free architecture. Finally, in Fig.~\ref{p6fig10}-Fig.~\ref{p6fig12}, we compare the small-cell and cell-free architecture relying on the proposed scheme. {{The figures are shown for $10^6$ Monte-Carlo runs, except for the BER plot, where it is $10^8$.}}
\par  

We consider a $500~m \times 500~m \times 120~m$ space for all simulation studies. The height is fixed to $120~m$, which is the maximum allowable height for UAVs without any permit in many of the countries \cite{mozaffari2019tutorial}.  The simulations are carried out at millimeter wave frequencies and the parameters are chosen from \cite{yan2019performance, nie201428, rao2019modeling, maccartney201473, rappaport2015wideband}, which are reproduced below in Table \ref{p6table:TxParams}. The path loss is governed by the equation $PL = 32.4+20\log d+20\log f$, where $d$ is the distance and $f$ is the carrier frequency in $GHz$~\cite{maccartney2017rural}. The transmitter antenna gain is fixed to $10\log M$, where $M$ is the total number of elements in the antenna \cite[(6-49)]{balanis2016antenna}.  
\begin{table}[h]
	\begin{center}
		\begin{tabular}{|l|r|}
			\hline
			Parameter & Value \\
			\hline 
			\hline 
			Frequency $(GHz)$& $73.5$ \\
			\hline
			Bandwidth $(GHz)$& $5$ \\
			\hline
			TxPower $(dBm)$& $14.6$ \\
			\hline
			Other Loss $(dB)$ & $12.7$ \\
			\hline 
			Rx Gain $(dBi)$ & $27$ \\
			\hline 
			Rx NF $(dBm)$& $7$ \\
			\hline 
			Rx Noise $(dB)$ & $-76.8$\\
			\hline
			Modulation  & $ 64 QAM $\\
			\hline
		\end{tabular}
	\end{center}
	\caption{Simulation parameters}
	\label{p6table:TxParams}
\end{table}

\par

We consider a spherical 3D cellular structure for evaluating the performance of the small-cell architecture. The BS's pair of transmit AAs have been placed at $120~m$ height separated by $120~m$ distance. Both these AAs have planar layout in the $x-y$ plane looking downwards. We considered both $15 \times 15$ and $30 \times 30$ AAs.  The radius of the micro-cell is varied from $15~m$ to $40~m$. The macro-cells are grouped in such a way that two cells sharing the resources are located at least a distance of four times the micro-cell radius. Hence, the number of interfering users is fixed based on the macro-cell radius.   The positions of desired user and interferers are randomly selected in the respective micro-cell.  {{However, the azimuth and elevation angles of one of the interferer with respect to $AA1$ is deliberately made within $2^0$ same as that of the desired user. This is done to highlight the advantage of the proposed scheme over the conventional 3D beamformer.}}  The performance measures are plotted against the cell radius in Fig~\ref{p6fig2} and Fig.~\ref{p6fig3}.  

\par 

Fig.~\ref{p6fig2} shows the VSE of the system, where the simulation results are compared against the theoretically obtained values. Explicitly, for estimating the theoretical VSE value, we evaluated (\ref{p6eqn15}) theoretically and the integral in (\ref{p6eqn16}) is evaluated through Monte Carlo runs for various user locations. We have to bear in mind that while estimating the average VSE, the volume should be multiplied by the reuse factor in order to account for the resource reuse. Observe that the values obtained through simulations and those from theory are closely matching. It can be seen that as the cell radius increases, the average VSE decreases. As a further dominant trend, the VSE is improved as the Rician factor ($K$) is increased and the general VSE trend versus these parameters are as expected. Similarly, when the number of elements in the BS AA is increased, this improves the interference rejection capability, hence improving the VSE. 
\par 
In Fig.~\ref{p6fig3}, the average spectral efficiency (SE) in $bps/Hz$ is plotted against the cell radius, which is calculated using (\ref{p6eqn16}). The theoretical values are compared against those obtained through the simulations and it can be seen that these values exhibits similar trends. In contrast to the VSE, the SE is increased with cell radius and it is improved both with $K$ and with the number of elements in the BS AAs. {{Note that we have approximated a Gaussian mixture distribution using a single Gaussian distribution to derive the theoretical values of VSE and SE. This approximation introduces the gap between the theory and simulations in both Fig.~\ref{p6fig2} and Fig.~\ref{p6fig3}. }}
%\begin{figure}
%	\centering
%	\includegraphics[width = 0.6\textwidth]{Figs/p6fig3.eps}
%	\caption{BER vs Cell Radius.}	
%	\label{p6fig4}
%\end{figure} 
\par 
%The average BER performance of the proposed scheme is portrayed in Fig. \ref{p6fig4}. For estimating the BER, we used $64$-level $QAM$ as the signal constellation and a maximum likelihood (ML) detector at the receiver. The average BER is improved when the cell radius, the Rician Factor $K$ and the number of elements in the BS antennas are increased. This is as expected, since in these cases, the effect of interference is reduced. 

Next, the performance of the proposed scheme quantified in the cell-free architecture is considered. The users are randomly located in the $500 \times 500 \times 120~m$ space specified. Note that here all users have access to all resources. The performance is studied for different number of access points (NA), number of elements per access points (NE) and number of users (NU).  

\par
\begin{figure}
	\centering
	\includegraphics[width = 0.45\textwidth]{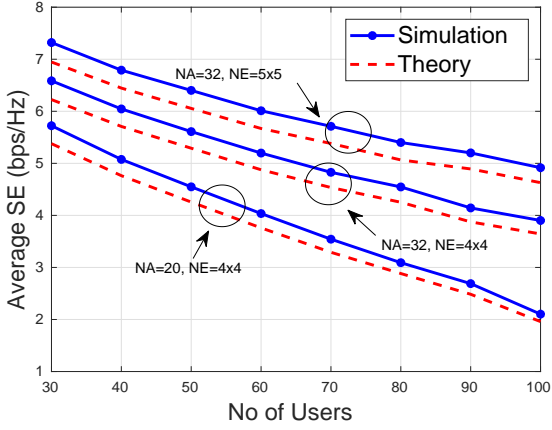}
	\caption{Average SE vs No of Users.}	
	\label{p6fig9}
\end{figure}

 
In Fig.~\ref{p6fig5}, the average VSE is plotted against the NA, whereas the corresponding average SE vs. NA is shown in Fig. ~\ref{p6fig6} for various Rician factors and NE.  It can be seen that both the average VSE and the average SE increases with the NA and NE. This is clear from the fact that as either NA or NE increases, the transmission nulls created towards the sources of interferences become deeper, thereby improving the SINR.  %Fig.~\ref{p6fig7} shows the average BER vs. NA for the Rician factor $K=10~dB$. The curves are shown for different NE and NU values. The average BER performance is improved upon increasing both NA and NE. The average BER is increased with NU. 
%          \begin{figure}
%          	\centering
%          	\includegraphics[width = 0.6\textwidth]{Figs/p6fig8.eps}
%          	\caption{Average BER vs No of APs.}	
%          	\label{p6fig7}
%          \end{figure} 
The average VSE and the average SE have been plotted vs. NU in Fig. ~\ref{p6fig8} and Fig.~\ref{p6fig9}, respectively for the Rician factor of $K=10~dB$. The curves are shown for different NA and NE. Observe that the average SE is reduced upon increasing NU, since the SINR is reduced as the number of users is increased. However, in contrast to the average SE, the average VSE initially increases with NU and beyond its maximum it reduces. This is because the reuse factor is increased with NU. Hence, even though the SINR is reduced with NU, the combined effect of the reuse factor and NU results in an increase of the average VSE upto a certain maximum. If however the NU is increased further, the SINR reduction becomes the dominant factor, thereby reducing the average VSE. {{Similar to the small-cell architecture, a gap can be seen between the theory and simulations in cell-free structure also, which is resulted from the single Gaussian approximation of Gaussian mixture distribution in deriving the theoretical expression.}}

\par 

Fig.~\ref{p6fig10}-Fig.~\ref{p6fig12} compare the performance of the cell-based and cell-free architectures in the context of the proposed 3D CBF technique vs. Rician factor $K$. We consider a pair of $16 \times 16$ element arrays for the small cell based structure. The radius of the macro-cell is fixed to $20~m$ and $30~m$, corresponding to 50 and 20 users, respectively. As for the cell-free architecture, the number of APs is fixed to 16, i.e., $NA=32$ so that 16 APs contributing to each of the CPU. Each AP is constituted by a $4 \times 4$-element array, i.e., $NE=4 \times 4$. It may be noted that the array structure of the small-cell configuration as well as of the $NA$ and $NE$ in the cell-free configuration are chosen in such a way that the number of elements per antenna is 256 in both cases. As in the case of the small-cell configuration, in cell-free architecture, the number of users ($NU$) is set to 50 and 20. Fig.~\ref{p6fig10}, Fig.~\ref{p6fig11} and Fig.~\ref{p6fig12} show the average VSE, average SE and average BER vs. the Rician factor, respectively. It can be seen that the performance of the small-cell based architecture is better than that of the cell-free configuration for all the three parameters. The performance difference between the two architectures can be explained by considering the specific antenna structures. In the small-cell configuration, the elements of each antenna are arranged using half-wave-length spacing. By contrast, for the cell-free configuration, the elements are distributed randomly in space, with separations more than half the wavelength. Hence, in the cell-free case, the beampattern is irregular and may have grating lobes, which decreases the SINR.
%Hence, the antennas in the small-cell based architecture are able to form sharper directional beams and thereby increasing the SINR and improving the performance. 
%Observe in Fig.~\ref{p6fig11} that the cell-based architecture has a better performance in terms of average SE. However, the average VSE of the cell-free scenario is much higher than that of the cell-based architecture. This is because maintaining a higher reuse factor becomes possible for the cell-free architecture. The two architectures are compared in terms of the average BER performance in Fig.~\ref{p6fig12}. The average BER of the small-cell architecture is better than that of the cell-free architecture for NA=30. However, for NA =60, the average BER performance of cell-free architecture is better than that of the cell-based one. This reinforces the trend of migrating towards a cell free architecture for 6G systems, as it leads to higher average VSE and better BER performance, especially for large NA values.

\par
\begin{figure}
	\centering
	\includegraphics[width = 0.45\textwidth]{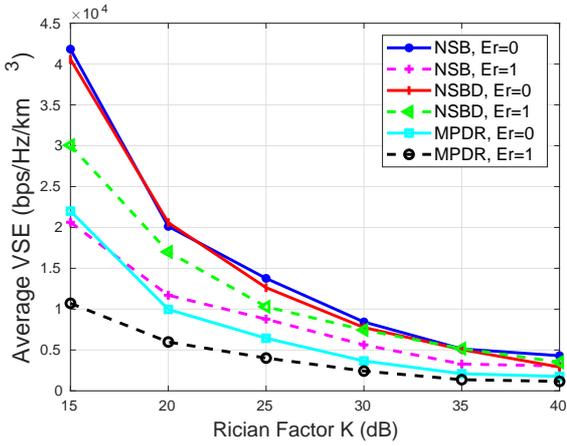}
	\caption{Comparison of different schemes with and without position errors in small-cell architecture. Er represents error variance in position estimation.}	
	\label{p6fig13}
\end{figure}

 
{{Fig.~\ref{p6fig13} compares the performance of various transmit precoding (TPC) schemes detailed in Section \ref{prop},viz. ZFP, ZFP-D and MPDR in small-cell scenarios in terms of the average VSE. It also shows the performance of the TPC schemes in the presence of position estimation errors. The errors in position estimation are introduced by adding zero mean complex Gaussian random variables with $1~m$ variance  to the actual locations of all users. These erroneous positions are used for deriving the TPC vector. It can be readily seen that among the three schemes, ZFP outperforms the other schemes, when there are no position estimation errors. However, the performance of ZFP-D is more robust against position errors than the other two schemes, where  there is less than $25\%$ degradation in average VSE even at $1~m$ position errors. The performance of MPDR is worst. This is because, the optimization of the MPDR solution aims for minimizing the interferences, while retaining a distortionless response in the desired direction. By contrast, the other two schemes try to introduce nulls in the known interferers' directions, which makes them more efficient in terms of interference rejection.}}
\par

\begin{figure}
	\centering
	\includegraphics[width = 0.45\textwidth]{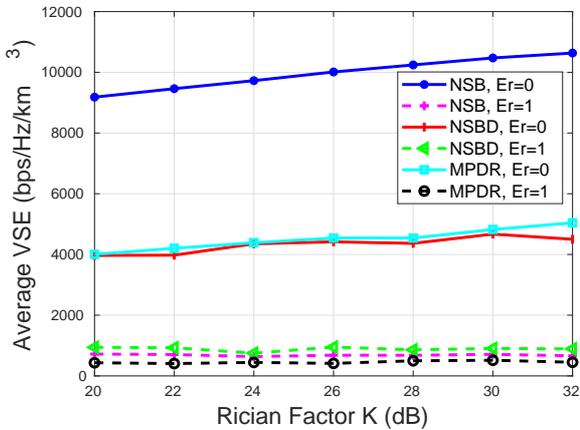}
	\caption{Comparison of different schemes with and without position errors in cell-free architecture. Er represents error variance in position estimation.}	
	\label{p6fig14}
\end{figure}

{{In Fig.~\ref{p6fig14}, the comparison of various schemes and the effect of position errors is shown for cell-free architecture. In contrast to the small-cell architecture, the performance is very sensitive to position errors and the performance degrades to less than $10\%$ with $1~m$ position errors. This trend is explained in the last paragraph of this section. }}

\par
\begin{figure}
	\centering
	\includegraphics[width = 0.45\textwidth]{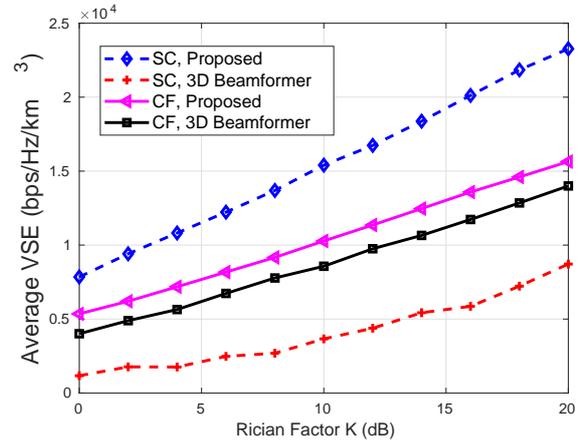}
	\caption{Comparison of proposed scheme and conventional 3D beamformer using single antenna array with zero-forcing precoding. }	
	\label{p6fig15}
\end{figure}

  
{{The proposed scheme is compared against the conventional 3D beamformer using a single antenna array for both the small-cell and the cell-free architecture  in Fig.~\ref{p6fig15}.   It can be seen that for the small-cell architecture, there is a large gap between the proposed and the conventional single antenna array based 3D beamformer, whereas in the cell-free architecture this gap is small. This trend is explained as follows.}}
\par 
{{In the cell-free architecture, the elements in each antenna array, except for those in the same APs, are randomly located in space at much more than half-wavelength spacing. By contrast, in the small-cell architecture, they are placed exactly at half-wavelength distance. Since, the frequency is inversely proportional to the wavelength, the antenna arrays of the small-cell architecture can be considered to be designed for much lower frequency. It may be noted that the beamwidth of an antenna array is proportional to the ratio of the operating wavelength and the element spacing \cite[pp. 47]{van2004optimum}. In other words, it is proportional to the ratio of the designed and the operating frequencies. In small-cell architectures, the system is operating at the designed frequency itself, whereas in cell-free architecture, it is operating at higher frequency, which makes the beamwidth very narrow in the cell-free case. Because of this very narrow beamwidth, even the scheme using a single antenna array is capable of separating two users, which are as close as 2 degrees in terms of their angular separation. However, this narrow beamwidth makes the performance highly sensitive to the position errors, which causes the degradation in performance in the face of position errors as seen in Fig. \ref{p6fig14}. The effect of position errors on the performance can be reduced by keeping more elements in the APs and thereby reducing the beamwidth. Note that, in this case, the performance gap between the proposed scheme and the conventional 3D beamformer using a single antenna array will further increase and our scheme will become better and better.} }

\section{Conclusions}
\label{p6con}

A novel CBF strategy has been proposed for supporting 3D  users, which is a 6G system design requirement. The proposed scheme is suitable for numerous applications, especially those related to UAVs and IoT. The proposed technique has been characterized both in small-cell and cell-free wireless topologies. We semi-analytically  evaluated the VSE and SE of the proposed scheme and compared them against our simulation results. Furthermore, we also evaluated the BER performance of the proposed scheme. Even though the simulations have been conducted for millimeter wave frequencies, the methodology is readily applicable for Terahertz (THz) frequencies. Furthermore, we compared the VSE of both the cell-free and the small-cell based architectures and observed that the small-cell based architectures attain better VSE. {{Also, it is less sensitive to position estimation errors.} }
\par 
Again, we have considered zero forcing JTPC relying on a 3D CBF scheme. An advantage of our PBTBF is that the JTPC vector can be easily computed based on the relative positions of UEs as well as BSs and it dispenses with the channel estimation requirements. Hence, this is eminently suitable for UAV based applications, where the UEs and BSs may be in perpetual movement. An interesting future work item may be to consider the mobility aspects of our scheme, where one has to consider the choice of a suitable modulation scheme, optimal movement trajectory to maximize the performance and so on. Another promising future work item is to incorporate the 3D CBF techniques in airplane aided communication systems, which has been proposed as a potential technology for improving the coverage probability of 6G systems \cite{Huang2019, 9491998}. In this case, the minimum safety distance between two airplanes, the cooperation strategy among them and  many other questions have to be addressed and will be a challenging problem to consider.

\appendices
\section{Proof of Theorem \ref{p6thm1}}
\label{p6proofthm1}
Under the conditions of Theorem \ref{p6thm1}, the average capacity per unit bandwidth can be expressed as:
\begin{align} 
\bar{C}_{h,u} = \int_{0}^{\infty} \log_2\left(1+\eta\right)f_{\eta}(\eta) d\eta,
\label{p6eqn14}
\end{align}
where $f_{\eta}(\eta)$ is the distribution of $\eta$ given in Lemma \ref{p6_lem_dist} of Appendix \ref{p6_app_d}.
\par
Let $c_{\sigma} = \frac{\sigma_s^2}{\sigma_I^2}$. Now, make a change of variable $z = \frac{c_{\sigma}}{\eta+c_{\sigma}}$ in (\ref{p6eqn13}) and substitute it into (\ref{p6eqn14}). Then we will get:
\begin{align}
\bar{C}_{h,u} &=\frac{\left(1+k_s^2\right)}{\log 2} I,
%	\bar{C}_{h,u} &=\left(1+k_s^2\right) \int_{0}^{1} \log_2\left(1+\frac{c_{\sigma}}{z}-c_{\sigma}\right)e^{-k_s^2z}\left(1 - \frac{k_s^2}{1+k_s^2}z\right)dz \nonumber\\
%&=\left(1+k_s^2\right) I,
%&=\left(1+k_s^2\right) \int_{0}^{1} \log_2\left(\frac{c_{\sigma}\left(1+\frac{\left(1-c_{\sigma}\right)}{c_{\sigma}}z\right)}{z}\right)e^{-k_s^2z}\left(1 - \frac{k_s^2}{1+k_s^2}z\right)dz  \nonumber \\
%&=\frac{1}{ \left(c_{\sigma}-1\right)k_s^2 \log 2} I,
\label{p6eqn15}
\end{align} 
where the integral $I$ is:
\begin{align}
I &= \int_{0}^{1} \log\left(1-c_{\sigma}+\frac{c_{\sigma}}{z}\right)\left(1 - \frac{k_s^2}{1+k_s^2}z\right)e^{-k_s^2z}dz \nonumber \\
&= I_1 - I_2 + I_3,
\label{p6eqn15a}
\end{align}

{{where the terms $I_1$, $I_2$ and $I_3$ are evaluated below:}}
\begin{align}
I_1 &= \log c_{\sigma} \int_{0}^{1} \left(1 - \frac{k_s^2}{1+k_s^2}z\right)e^{-k_s^2z}dz \nonumber \\
&= \frac{1}{1+k_s^2}\log c_{\sigma}.
\label{p6eqn24}
\end{align}
{{(\ref{p6eqn24}) is obtained using integration by parts as: }}
\begin{align}
I_2 &= \int_{0}^{1}  \left(1 - \frac{k_s^2}{1+k_s^2}z\right)\log z e^{-k_s^2z}dz \nonumber \\
&=  \left(I_{21} - I_{22}\right),
\label{p6eqn25}
\end{align}
\textbf{{where $I_{21}$ is evaluated using \cite[(4.331)]{gradshteyn2014table} and \cite[(8.212)]{gradshteyn2014table} as:}}
\begin{align}
I_{21} &=\int_{0}^{1}  \log z e^{-k_s^2z}dz =\frac{1}{k_s^2}\int_{0}^{1} \frac{e^{-k_s^2z}-1}{z}dz \nonumber \\
&= \frac{1}{k_s^2} \left(Ei(-k_s^2) - \log k_s^2 - \zeta \right).
\label{p6eqn27}
\end{align}
{{Now $I_{22}$ is obtained using integration by parts considering $z\log z$ as the first term and is given below:}}
\begin{align}
I_{22} &= \frac{k_s^2}{1+k_s^2}\int_{0}^{1} z\log z e^{-k_s^2z}dz \nonumber \\
&=\frac{1}{k_s^2\left(1+k_s^2\right)}\left(1 - e^{-k_s^2}\right) + \frac{1}{1+k_s^2}I_{21}. 
\label{p6eqn26}
\end{align}
Hence, 
\begin{align}
I_2 = \frac{1}{\left(1+k_s^2\right)}\left(Ei(-k_s^2) - \log k_s^2 - \zeta \right)-\frac{\left(1 - e^{-k_s^2}\right)}{k_s^2\left(1+k_s^2\right)}.
\label{p6eqn28}
\end{align}
{{Now, $I_3$ can be found out as follows:}}
\begin{footnotesize}
	\begin{align}
	I_3 &= \int_{0}^{1} \log\left(1+\frac{\left(1-c_{\sigma}\right)z}{c_{\sigma}}\right)\left(1 - \frac{k_s^2}{1+k_s^2}z\right)e^{-k_s^2z}dz \nonumber \\
	&= \int_{0}^{1}\sum_{n=1}^{\infty} \frac{(-1)^{n-1}}{n}\left(\frac{\left(1-c_{\sigma}\right)z}{c_{\sigma}}\right)^n\left(1 - \frac{k_s^2}{1+k_s^2}z\right)e^{-k_s^2z}dz \nonumber \\
	&= \sum_{n=1}^{\infty}\frac{1}{n}(-1)^{n-1}\left(\frac{1-c_{\sigma}}{c_{\sigma}}\right)^n \left(I_{31} - \frac{k_s^2}{1+k_s^2}I_{32}\right).
	\label{p6eqn29}
	\end{align}
\end{footnotesize}
{{(\ref{p6eqn29}) is obtained changing the order of summation and integration and rearranging the terms. Now, the integrals $I_{31}$ is obtained using \cite[(3.351)]{gradshteyn2014table} as:}}
\begin{align}
I_{31} &= \int_{0}^{1} z^ne^{-k_s^2z} dz \nonumber \\
&= \left(k_s^2\right)^{-(n+1)}\gamma\left(n+1,k_s^2\right),
\label{p6eqn30}
\end{align}
{{where $\gamma(.,.)$ is the incomplete Gamma function. Similarly $I_{32}$ can also be obtained. Now combining $I_{31}$ and $I_{32}$ and using the relation $\gamma\left(s+1,x\right)=s\gamma(s,x)-x^se^{-x}$ \cite[(8.356(1))]{gradshteyn2014table}, $I_3$ can be simplified as:}}
\begin{footnotesize}
	\begin{align}
	I_3
	&= \sum_{n=1}^{\infty}\frac{1}{n}(-1)^{n-1}\left(\frac{1-c_{\sigma}}{c_{\sigma}}\right)^n \left(k_s^2\right)^{-(n+1)}\left(\gamma\left(n+1,k_s^2\right) \right. \nonumber \\
	&~~~~~~\left. - \frac{n+1}{1+k_s^2} \gamma\left(n+1,k_s^2\right) + \frac{\left(k_s^2\right)^{n+1}}{1+k_s^2}e^{-k_s^2}\right) \nonumber \\
	&= \frac{1}{1+k_s^2} \sum_{n=1}^{\infty}\frac{(-1)^{n-1}}{n}\frac{\left(k_s^2-n\right)}{\left(k_s^2\right)^{(n+1)}}\left(\frac{1-c_{\sigma}}{c_{\sigma}}\right)^n \gamma\left(n+1,k_s^2\right) \nonumber \\
	&~~~~~~~~~~~~~~~~~-\frac{e^{-k_s^2}}{1+k_s^2} \log \left(c_{\sigma}\right)\nonumber \\
	&=\frac{1}{1+k_s^2}\sum_{n=1}^{\infty}(-1)^{n}\left(\frac{1-c_{\sigma}}{c_{\sigma}}\right)^n \left(k_s^2\right)^{-(n+1)}\gamma\left(n+1,k_s^2\right)  \nonumber \\
	&~+ \frac{1}{1+k_s^2}\sum_{n=1}^{\infty}(-1)^{n-1}\left(\frac{1-c_{\sigma}}{c_{\sigma}}\right)^n \left(k_s^2\right)^{-n}\gamma\left(n,k_s^2\right) \nonumber \\
	&~ + \frac{1}{1+k_s^2}\sum_{n=1}^{\infty}(-1)^{n}\frac{1}{n}\left(\frac{1-c_{\sigma}}{c_{\sigma}}\right)^n e^{-k_s^2} -\frac{e^{-k_s^2}}{1+k_s^2} \log \left(c_{\sigma}\right)\nonumber \\
	&= \frac{1}{1+k_s^2}\sum_{n=1}^{\infty}\left(\frac{c_{\sigma}-1}{c_{\sigma}k_s^2}\right)^n \left(\frac{\gamma\left(n+1,k_s^2\right)}{k_s^2} -\gamma\left(n,k_s^2\right) \right). 
	\label{p6eqn31}
	\end{align}
\end{footnotesize}
{{Expanding the infinite summation in (\ref{p6eqn31}) (represented by $I_{T1}$) and combining common terms, we will get:}}
\begin{align}
I_{T1} &= \left(\frac{c_{\sigma}-1}{c_{\sigma}k_s^2}\right)\left[ \gamma(2,k_s^2) - \gamma(1,k_s^2)\right] \nonumber \\
&+ \left(\frac{c_{\sigma}-1}{c_{\sigma}k_s^2}\right)^2\left[ \left(k_s^2\right)^{-1}\gamma(3,k_s^2) - \gamma(2,k_s^2)\right] + ... \nonumber \\
%&= \sum_{n=1}^{\infty}(-1)^{(n-1)}\left(\frac{1-c_{\sigma}}{c_{\sigma}}\right)^n \left(k_s^2\right)^{-(n+1)} \left(1+\frac{1-c_{\sigma}}{c_{\sigma}}\right)\gamma\left(n+1,k_s^2\right) + \left(k_s^2\right)^{-1}\left(\frac{1-c_{\sigma}}{c_{\sigma}}\right)\gamma(1,k_s^2)  \nonumber \\
&= \frac{1}{c_{\sigma}}\sum_{n=1}^{\infty}\left(\frac{c_{\sigma}-1}{c_{\sigma}}\right)^n \left(k_s^2\right)^{-(n+1)} \gamma\left(n+1,k_s^2\right) \nonumber \\
&~~~~~~~~ + \left(k_s^2\right)^{-1}\left(\frac{1-c_{\sigma}}{c_{\sigma}}\right)(1 - e^{-k_s^2}).
\label{p6eqn33}
\end{align}
{{We used the relation $\gamma(1,k_s^2) = (1 - e^{-k_s^2})$ \cite[(8.352)]{gradshteyn2014table} for obtaining the last term of (\ref{p6eqn33}). Now for all $z > 0$, i.e. $k_s^2$, $\gamma(n,k_s^2) =\frac{1}{n}\left(k_s^2\right)^n{}_1F_1\left(n,n+1,-k_s^2\right) = \frac{1}{n}\left(k_s^2\right)^n\sum_{p=0}^{\infty}\frac{(n+1)_p}{(n+2)_p}\frac{(-k_s^2)^p}{p!}$ \cite[(4.4.6)]{rota1976encyclopedia}.   Hence, the first term in RHS of (\ref{p6eqn33}) (represented by $I_{T2}$) can be written as:}}	

\begin{align}
I_{T2} &=\frac{1}{c_{\sigma}}\sum_{n=1}^{\infty}\sum_{p=0}^{\infty} \frac{(n+1)_p}{(n+2)_p} \frac{(-k_s^2)^p}{p!} \frac{1}{n+1}\left(\frac{c_{\sigma}-1}{c_{\sigma}}\right)^{n} \nonumber \\
&=\frac{1}{c_{\sigma}}\sum_{n=0}^{\infty}\sum_{p=0}^{\infty} \frac{(n+1)_p}{(n+2)_p} \frac{(-k_s^2)^p}{p!} \frac{1}{n+1}\left(\frac{c_{\sigma}-1}{c_{\sigma}}\right)^{n} \nonumber \\ &~~~~~~~~~~~~~-\frac{1}{c_{\sigma}}\sum_{p=0}^{\infty} \frac{(1)_p}{(2)_p} \frac{(-k_s^2)^p}{p!} \nonumber \\
&=\frac{1}{c_{\sigma}}\sum_{n=0}^{\infty}\sum_{p=0}^{\infty} \frac{(n+1)_p}{(n+2)_p} \frac{n!}{n+1}\frac{\left(\frac{c_{\sigma}-1}{c_{\sigma}}\right)^{n}}{n!} \frac{(-k_s^2)^p}{p!} \nonumber \\ &~~~~~~~~~~~~~-\frac{1}{c_{\sigma}k_s^2}\left(1-e^{-k_s^2}\right) \nonumber \\
&=\frac{1}{c_{\sigma}}\sum_{n=0}^{\infty}\sum_{p=0}^{\infty} \frac{1}{n+p+1}(1)_n \frac{\left(\frac{c_{\sigma}-1}{c_{\sigma}}\right)^{n}}{n!} \frac{(-k_s^2)^p}{p!} \nonumber \\ &~~~~~~~~~~~~~-\frac{1}{c_{\sigma}k_s^2}\left(1-e^{-k_s^2}\right) \nonumber \\
&=\frac{1}{c_{\sigma}}\sum_{n=0}^{\infty}\sum_{p=0}^{\infty} \frac{(1)_{n+p}(1)_n}{(2)_{n+p+1}} \frac{\left(\frac{c_{\sigma}-1}{c_{\sigma}}\right)^{n}}{n!} \frac{(-k_s^2)^p}{p!} \nonumber \\ &~~~~~~~~~~~~~-\frac{1}{c_{\sigma}k_s^2}\left(1-e^{-k_s^2}\right) \nonumber \\
&=\frac{1}{c_{\sigma}}\phi_1\left(1,1,2;\frac{c_{\sigma}-1}{c_{\sigma}},-k_s^2 \right)-\frac{1}{c_{\sigma}k_s^2}\left(1-e^{-k_s^2}\right),
\label{p6eqn34}
\end{align}
{{where $\phi_1(.,.,.;.,.)$ is the Confluent Hypergeometric function of two variables \cite[(A.38, pp. 212)]{mathai2009h}. Now using \cite[(Eq. 8)]{kemp2013new} and further exploiting \cite[(3.352(5), 3.352(6)), pp. 341]{gradshteyn2014table}, one can write:}}
\begin{footnotesize}
	\begin{align}
	&\phi_1\left(1,1,2;\frac{c_{\sigma}-1}{c_{\sigma}},-k_s^2 \right) = \int_{0}^{1}\left(1 - u\frac{c_{\sigma}-1}{c_{\sigma}}\right)^{-1}e^{-k_s^2u}du  \nonumber \\
	&~~~=\frac{c_{\sigma}}{c_{\sigma}-1}e^{-\frac{c_{\sigma}k_s^2}{c_{\sigma}-1}}\left(Ei\left(\frac{c_{\sigma}k_s^2}{c_{\sigma}-1}\right) - Ei\left(\frac{k_s^2}{c_{\sigma}-1}\right)\right).
	\label{p6eqn35}
	\end{align}		
\end{footnotesize}

{{ Finally combining (\ref{p6eqn24}), (\ref{p6eqn28}), (\ref{p6eqn31}), (\ref{p6eqn33}), (\ref{p6eqn34}) and (\ref{p6eqn35}) and substituting in (\ref{p6eqn15}) will result in (\ref{p6eqn15e}).}}

\section{Proof of Theorem \ref{p6thm2}}
\label{p6proofthm2}
{{When selected constellation $s(m), m \in \{1, 2, ..., M\}$ is symmetrically spaced around zero, $\sum_{m=1}^{1}s(m) =0$ and since they are equally likely, $p(m)=\frac{1}{M},~\forall~m$. Hence, $\mu(m)$ in (\ref{p6eqn19}) is zero, which make the approximated complex Gaussian distribution for both the signal, i.e. $x_s$ and the interference plus noise, i.e., $x_I$ zero mean. Therefore, both $\|x_s\|^2$ and $\|x_I\|^2$ are exponentially distributed with parameters $\frac{1}{\sigma_s^2}$ and $\frac{1}{\sigma_I^2}$, respectively. Hence, the distribution of $\eta = \frac{\|x_s\|^2}{\|x_I\|^2}$ is given below:}}
\begin{align}
f_{\eta}(\eta) = \frac{c_{\sigma}}{\left(\eta + c_{\sigma}\right)^2},~\eta>0,
\label{p6eqn18a}
\end{align}
where $c_{\sigma} = \frac{\sigma_s^2}{\sigma_I^2}$. Hence, 
\begin{align} 
\bar{C}_{h,u} &= \int_{0}^{\infty} \log_2\left(1+\eta\right)\frac{c_{\sigma}}{\left(\eta + c_{\sigma}\right)^2} d\eta \nonumber \\
&\overset{(a)}{=}\frac{1}{\log 2}\left(\left[-\log\left(1+\eta\right)\frac{c_{\sigma}}{\eta+ c_{\sigma}}\right]_0^{\infty} \right. \nonumber \\
&~~~~\left.+ \int_{0}^{\infty} \frac{c_{\sigma}}{(\eta+1)(\eta+c_{\sigma})}d\eta \right) \nonumber \\
& = \frac{c_{\sigma}}{c_{\sigma}-1}\left[\log_2 \left(\frac{\eta +1 }{\eta + c_{\sigma}}\right) \right]_0^{\infty},
\label{p6eqn18b}
\end{align}
{{where (a) is obtained using integration by parts. The first term of RHS of (a) is zero, which can be obtained using L'Hosptials' rule \cite{kreyszig2008advanced}. Finally, evaluating the limits in (\ref{p6eqn18b})  will result in (\ref{p6eqn15f}), which proves the Theorem.}}
\section{Distribution of $\eta$ }
\label{p6_app_d}
\begin{lem}
	\label{p6_lem_dist}
	{{	Let the transmitted symbols $s_n,~\forall~n \in \left\{0, 1, ..., N\right\}$ be chosen from an $M$-ary constellation $s(m), m \in \left\{0, 1, ..., M-1\right\}$ having probability $p(m)$. Then the distribution of $\eta$ given in (\ref{p6eqn8}) can be approximated as:}}
	\begin{align}
	f_{\eta}(\eta) =\tilde{k}  e^{\frac{\eta k_s^2}{\eta+\frac{\sigma_s^2}{\sigma_I^2}}}\left(\frac{\eta+\frac{\sigma_s^2}{\sigma_I^2\left(1+k_s^2\right)}}{\left(\eta+\frac{\sigma_s^2}{\sigma_I^2}\right)^3} \right),~\eta>0,
	\label{p6eqn18}
	\end{align}
	{{	where $\tilde{k}=e^{-k_s^2}\left(1+k_s^2\right)\frac{\sigma_s^2}{\sigma_I^2}$ is a constant with $k_s = \frac{|\mu_s|}{\sigma_s}$. Here, $\mu_s$ and $\sigma_s^2$ are the mean and variance of signal component, respectively and $\sigma_I^2$ is the variance of interference plus noise component and are given by:}}
	
	\begin{align}
	\mu_s = \sum_{m=0}^{M-1}p(m)\mu(m),
	\label{p6eqn19}
	\end{align}
	where $\mu(m) = \sqrt{\frac{P_rK}{K+1}}\left(\e_{L0}+\e_{R0}\right)^H\e_0s(m)$.
	\begin{align}
	\sigma_s^2 =\sum_{m=0}^{M-1}p(m)\left(\frac{2P_r|s(m)|^2}{K+1}\|\e_0\|^2 +\left|\mu(m) - \mu_s\right|^2 \right).
	\label{p6eqn20}
	\end{align}
	\begin{align}
	\sigma_I^2 = \sum_{m=0}^{M-1}p(m)\frac{2P_r|s(m)|^2}{K+1}\left\|\sum_{n=1}^{N}\e_n\right\|^2+\sigma^2.
	\label{p6eqn21}
	\end{align}
	\begin{proof}
		\label{p6_lem_dist_proof}
		{{	 The signal and interference plus noise components are }}
		\begin{align}
		x_s = \sqrt{P_r}\h_0\e_0s_0
		\label{p6eqn21a}
		\end{align}
		and 
		\begin{align}
		x_I = \sqrt{P_r}\h_0\sum_{n=1}^{N}\e_ns_n + n_0,
		\label{p6eqn21b}
		\end{align}
		{{  respectively. Since both $\h_L$ in (\ref{p6eqn3}) and $\h_R$ in (\ref{p6eqn4}) are distributed as $\CN(0,\I)$, we have $\h_L+ \h_R \sim \CN(0,2\I)$. Now, note that,}}
		\begin{align}
		\left(\e_{L0}+\e_{R0}\right)^H\e_i \approx 0, \forall~i \ne 0,
		\label{p6eqn10}
		\end{align}
		{{	which is a direct consequence of zero-forcing precoding. Hence, it can be deduced that for a given $s_n = \tilde{s}_n,~n \in \{0, 1, ..., N\}$, $x_s$ and $x_I$ are distributed as $\CN(\tilde{\mu}_s,\tilde{\sigma}_s^2)$ and $\CN(0,\tilde{\sigma}_I^2)$, respectively, where $\tilde{\mu}_s = \sqrt{\frac{P_rK}{K+1}}\left(\e_{L0}+\e_{R0}\right)^H\e_0\tilde{s}_0$, $\tilde{\sigma}_s^2 = \frac{2P_r|\tilde{s}_0|^2}{K+1}\|\e_0\|^2$ and $\tilde{\sigma}_I^2 =\frac{2P_r}{K+1}\|\sum_{n=1}^{N}\e_n\tilde{s}_n\|^2+\sigma^2$. }}
		\par 
		{{Note that the transmitted symbols are chosen from an $M$-ary constellation, where the probability of occurrence of $m^{th}$ symbol is $p(m)$. Hence, the unconditional distribution of $x_s$ and $x_I$ are mixture of complex Gaussian and are $\sum_{m=1}^{M}p(m)\CN(\tilde{\mu}_s(m),\tilde{\sigma}_s^2(m))$ and $\sum_{m=1}^{M}p(m)\CN(0,\tilde{\sigma}_I^2(m))$, respectively. Here,  $\tilde{\mu}_s(m) = \sqrt{\frac{P_rK}{K+1}}\left(\e_{L0}+\e_{R0}\right)^H\e_0s(m)$, $\tilde{\sigma}_s^2(m) = \frac{2P_r|s(m)|^2}{K+1}\|\e_0\|^2$ and $\tilde{\sigma}_I^2 =\frac{2P_r}{K+1}\|\sum_{n=1}^{N}\e_ns(m)\|^2+\sigma^2$.
				Now, for computational tractability, we will approximate these mixture distributions by a pair of complex Gaussian distributions,  having the same first-order and second-order moments. It can be deduced that the mean and variance of approximated complex Gaussian distribution of $x_s$ are given by (\ref{p6eqn19}) and (\ref{p6eqn20}), respectively and $x_I$ is a zero mean random variable with variance given by (\ref{p6eqn21}) \cite{gaussApprox}. }}
		
		\par 
		{{Now, based on the approximated complex Gaussian distributions of $x_s$ and $x_I$, the distribution of $x = \left|\frac{x_s}{x_I}\right|$ can be obtained from \cite[(7)]{nadarajah2018note} as:}}
		\begin{align}
		f_x(x) = \frac{2xe^{-k_s^2}}{\sigma_s^2 \sigma_I^2 c_x^2}~ {}_1 F_1\left(2,1;\frac{x^2k_s^2}{\sigma_s^2c_x}\right),~x>0,
		\label{p6eqn11}
		\end{align}
		where  we have $c_x =\frac{x^2}{\sigma_s^2} + \frac{1}{\sigma_I^2}$. Hence, $\eta=x^2, x>0$ is distributed as $f_{\eta}(\eta) = \frac{1}{2\sqrt{\eta}}f_x(\sqrt{\eta})$ and is given by:
		\begin{align}
		f_{\eta}(\eta) = \frac{e^{-k_s^2}}{\sigma_s^2 \sigma_I^2 c_{\eta}^2}~ {}_1 F_1\left(2,1;\frac{\eta k_s^2}{\sigma_s^2c_{\eta}}\right),~\eta>0,
		\label{p6eqn12}
		\end{align}
		where $c_{\eta} = \frac{\eta}{\sigma_s^2} + \frac{1}{\sigma_I^2}$. However using \cite{wolfram6}, we have ${}_1F_1(2,1;z)=e^z(1+z)$. Therefore, (\ref{p6eqn12}) becomes:
		\begin{align}
		f_{\eta}(\eta) = \frac{e^{-k_s^2}}{\sigma_s^2 \sigma_I^2 } \frac{e^{\frac{\eta k_s^2}{\sigma_s^2c_{\eta}}}}{c_{\eta}^2}\left(1+ \frac{\eta k_s^2}{\sigma_s^2c_{\eta}}\right),
		\label{p6eqn13}
		\end{align}
		which will reduces to (\ref{p6eqn18}).
	\end{proof}
\end{lem}
%\Urlmuskip=0mu plus 1mu
\bibliography{3dfoc} 
\bibliographystyle{ieeetr} 

\end{document}